\documentclass[conference]{IEEEtran}
\usepackage{amsmath,amssymb,latexsym,hhline,graphicx,enumerate,cite}

\makeatletter
\def\ps@headings{%
\def\@oddhead{\mbox{}\scriptsize\rightmark \hfil \thepage}%
\def\@evenhead{\scriptsize\thepage \hfil \leftmark\mbox{}}%
\def\@oddfoot{}%
\def\@evenfoot{}}
\makeatother 

\newcommand{\nchoosek}[2]{\left(\begin{array}{c}#1\\#2\end{array}\right)}

\newtheorem{theo}{Theorem}
\newtheorem{prop}{Proposition}
\newtheorem{lem}{Lemma}
\newtheorem{corol}{Corollary}

\begin{document}

\title{A Probability Model for Lifetime of Wireless Sensor Networks}

\author{\authorblockN{Moslem Noori and
Masoud Ardakani\\} 
\authorblockA{Department of Electrical and Computer Engineering,
University of Alberta, CANADA\\
Email: \{moslem, ardakani\}@ece.ualberta.ca}}

\maketitle
\begin{abstract}
Considering a wireless sensor network whose nodes are distributed
randomly over a given area, a probability model for the network
lifetime is provided. Using this model and assuming that packet
generation follows a Poisson distribution, an analytical expression
for the complementary cumulative density function (ccdf) of the
lifetime is obtained. Using this ccdf, one can accurately find the
probability that the network achieves a given lifetime. It is also
shown that when the number of sensors, $N$, is large, with an error
exponentially decaying with $N$, one can predict whether or not a
certain lifetime can be achieved. The results of this work are
obtained for both multi-hop and single-hop wireless sensor networks
and are verified with computer simulation. The approaches of this
paper are shown to be applicable to other packet generation models
and the effect of the area shape is also investigated.
\end{abstract}

\section{Introduction} \label{Section Introduction}
Wireless sensor networks (WSNs) are consisted of a set of cheap and
usually battery-powered devices, called sensors. Sensor limited
power usually necessitates a compromise between lifetime and other
parameters such as the data rate or the quality of the received
signal in the sink. It is usually impracticable to replace the
sensors batteries after their operation period. Hence, estimating
the network lifetime according to the initial energy in sensors is
essential for network design. According to such lifetime estimation,
one can choose the network parameters such as node density, data
rate and initial energy of the sensors to achieve the desired
lifetime.

Lifetime analysis has been studied in the literature based on
different definitions such as the number of dead nodes in the
network, network coverage and network
connectivity\cite{Chandr_Upperbound_ICC,Chandr_Upperbound_Role_Infocom,
Azad_Multipe_Globecom,Santi_Lifetime_cellbased,
Hou_Alpha_lifetime,Rai_Lifetime_Modeling_DATE}. Authors in
\cite{Chandr_Upperbound_ICC} derive an upper bound on the network
lifetime considering the spatial behavior of the data source. To
achieve this goal, they first consider a simplified version where
the data source is a specific point, and the source is connected to
the sink with a straight line consisting of relaying sensors. They
derive the optimum length of a hop and consequently the number of
hops in the path to minimize the total energy consumed for the data
delivery. Then, they remove the assumption of a source concentrated
on a point and assume that the source is distributed over an area.

In \cite{Chandr_Upperbound_Role_Infocom}, the results of
\cite{Chandr_Upperbound_ICC} are extended to the networks whose
nodes may perform different tasks of sensing, relaying and
aggregating. The results of \cite{Chandr_Upperbound_ICC} are also
extended to multiple-sink networks in \cite{Azad_Multipe_Globecom}.

Work reported in \cite{Santi_Lifetime_cellbased} studies the network
lifetime for a cell based network. It is assumed that $N$ nodes are
deployed over a hypercube. For the aim of energy conserving, the
area is divided to $n$ hypercubes (cells). Using occupancy theory
\cite{Kolchin_Random_allocation}, the distribution of the minimum
number of sensors within each cell is investigated when
$N,n\rightarrow \infty$. Then, authors study the lifetime for the
case when network remains almost surely connected. Using the number
of sensors in each cell, the network lifetime is lower bounded based
on the given lifetime of each sensor.

A lifetime study based on the area coverage is presented in
\cite{Hou_Alpha_lifetime}. It is assumed that the nodes have a
circular sensing region and are distributed over a squared area.
Using the stochastic geometry, theory of coverage process, and
assuming the size of the area goes to infinity, an expression for
the node density is derived to guarantee a $k$-coverage in the area.
It is shown that using the proposed density, the network lifetime is
upper bounded by $kT$ where $T$ is the given lifetime of each
sensor. Although the upper bound is derived for an asymptotic
situation when the area goes to infinity, it is shown through
simulation that the derived bound is also reasonable for networks
over a finite area.

Authors in \cite{Rai_Lifetime_Modeling_DATE} divide linear or
circular networks to some bins where each bin contains a
deterministically assigned number of nodes. The nodes within each
bin, however, are deployed randomly. Also, the lifetime is defined
as the time when a hole occurs in the routing scheme (i.e. death of
a bin). Assuming a fixed transmission power for each packet and
using the theory of stochastic processes, authors have found the
probability distribution function (pdf) of the network lifetime. In
addition, they propose a method to assign the number of nodes within
each bin in order to maximize the network lifetime.


It is worthy to note that other studies in the literature are
performed on the lifetime, e.g.
\cite{DuarteMelo_Hetereogen_Globecom,Ganz_Lifetime_LargNet_Elsevier,Bydere_reliable_IEICE,
Chen_Lifetime_Commletter,Coleri_Automata_WSNA}. However, the most
related ones to this work are those that we discussed earlier.

In this paper, we find the probability of reaching a certain
lifetime for randomly distributed networks based on the power
dissipation model of the sensors. More specifically, unlike
\cite{Santi_Lifetime_cellbased, Hou_Alpha_lifetime}, we do not
assume that the lifetime of a sensor is given in order to find the
network lifetime. Instead, we find the lifetime of a sensor (as a
random variable) based on its power dissipation and packet
generation model. Also, our analysis does not assume an infinite
area and infinite number of sensors. In comparison to
\cite{Rai_Lifetime_Modeling_DATE}, we consider totally randomly
deployed networks over more variant area shapes. In addition, both
fixed and adjustable transmission power are studied in this work.
Also, the definition of lifetime in our work is more general and can
include the case studied in \cite{Rai_Lifetime_Modeling_DATE} (to be
discussed in Section \ref{Section lifetime pdf multi-hop}).

Considering the randomness in packet generation and sensor
deployment in the area, the lifetime of a network is a random
variable. For a lifetime analysis of the network, it is needed to
have a knowledge of the lifetime of each individual sensor. In this
work, instead of assuming that the lifetime of each sensor is given
beforehand, we first perform a lifetime analysis at the sensor
level. To this end, we model the lifetime of a sensor as a random
variable and find its distribution based on the traffic model and
the power dissipation model in the sensor. Using this probabilistic
model of a sensor lifetime and the distribution of the sensors over
the area, the complementary cumulative distribution function (ccdf)
of the lifetime of a single-hop network is derived. From this ccdf,
the probability distribution function (pdf) of the lifetime is also
obtained. The single-hop analysis will be the base of our further
extensions.

In the proposed analysis, no asymptotic assumption is made on the
number of nodes. Nevertheless, an asymptotic analysis is provided,
which---with an error exponentially decaying with the number of
sensors---predicts whether or not a desired lifetime can be
achieved.

The above analysis is then extended to multi-hop networks. Since the
lifetime of the multi-hop networks is dependent on the routing
scheme, we study the lifetime ccdf under the \emph{maximum-lifetime}
\cite{Al-Karaki_Routing_Magazine2004} routing.

The methodologies of this work are applicable to more general
scenarios, some of them are discussed in this paper. For example, we
extend the results to different traffic models; to the case where
different sensors may have different initial energy or traffic load;
and to various area shapes.

The organization of this paper is as follows. In Section
\ref{Section Model} we introduce the system model and provide the
required definitions and assumptions. The lifetime analysis for
single-hop networks is studied in Section \ref{Section Prob Model
Single Hop}. Section \ref{Section lifetime pdf multi-hop} discuses
the lifetime analysis of multi-hop networks. Extensions to other
scenarios are discussed in Section \ref{Section Extnesion} and the
accuracy of the proposed method is verified through simulations in
Section \ref{Section Simulations}. The paper is concluded in Section
\ref{Section Conclusion}.

\section{System Model}\label{Section Model}
In this section, the components of the system model such as lifetime
definition, energy consumption model and network traffic model are
introduced.

\subsection{Lifetime Definition}
As mentioned, lifetime has a great significance in the design of
WSNs. Conceptually, lifetime means the time duration that the
network is operational and can perform its assigned task. Since
there is no unique measure of the network failure, the definition of
the lifetime is application-related.

In \cite{Heinzelman_TCOM, Madan_Distributed_Routing_TWireless,
Chang_Energy_Conserving_Infocom} lifetime is stated as the time when
the first node dies. Usually the remaining sensors in the network
can accomplish the network's assigned task. Therefore, another
definition based on the ratio of dead nodes to the total number of
nodes in the network is often used (e.g.
\cite{DuarteMelo_Hetereogen_Globecom
,Heinemann_ACM_Mobicom2001,Oym_ICC_2004}). Notice that this
definition includes the definition of lifetime based on the death of
the first node and therefore is more general. Other definitions
based on the communication connectivity or the coverage of the area
are also proposed for the lifetime
\cite{Hou_Alpha_lifetime,Santi_Lifetime_cellbased}.

In this study, we consider the network lifetime based on the ratio
of dead nodes to the total number of nodes, $\beta$. For multi-hop
networks, where the nodes close to the sink have more traffic load
than other nodes and die sooner, we will modify this definition.

\subsection{Energy Consumption Model}
The network lifetime is directly related to the sensors lifetime
and in other words the energy dissipated in the sensor nodes. The
consumed energy in sensors includes the energy required for sensing,
receiving, transmitting and processing of data. The total consumed
energy is usually dominated by the required energy for data
transmission.

Two cases may be considered for the transmission mode of the nodes
in the network. In the first case, nodes transmit with a fixed
transmission power. This usually results in a fixed transmission
range. In the second case, nodes use a mechanism to adjust their
transmission power based on their distance to the next hop or the
sink. Hence, the required energy for a packet transmission in sensor
$i$ can be modeled as \cite{Hein_Tranmission_model_2000}
\begin{align}
e(d_i)&= l(e_t d_i ^{\alpha} + e_o) \nonumber\\
&=kd_i^{\alpha}+c  \label{Power dissipation model}
\end{align}
where $l$ represents the packet length in bits, $d_i$ denotes the
distance between sensor $i$ and the next hop, $\alpha$ represents
the path loss exponent, $e_t$ shows the loss coefficient related to
1 bit transmission and $e_o$ is the overhead energy due to the
sensing, receiving and processing for the same amount of data. Also,
$k = le_t$ and $c = le_o$ represent the loss coefficient and the
overhead energy for a packet transmission respectively. The path
loss exponent depends on the local terrain and is determined by
empirical measurements. The typical value of $\alpha$ for WSNs is
from $2$ to $4$ \cite{Oym_ICC_2004}.

While this work is more focused on the transmission model
(\ref{Power dissipation model}), fixed transmission power is also
discussed.

\subsection{Traffic Model}
The traffic model of the network depends on the network application
and the behavior of sensed events. The data reporting process in
WSNs is usually classified into three categories: event-driven,
time-driven and query-driven \cite{Al-Karaki_Routing_Magazine2004}.
In the time-driven case, sensors send their data periodically to the
sink. Event-driven networks are used when it is desired to inform
the data sink about the occurrence of an event. In query-driven
networks, sink sends a request of data gathering when needed. In
this paper, our main focus will be on the event-driven networks with
Poisson model for packet generation.

Suppose that the events are independent (both temporally and
spatially) and occur with equal probability over the area. In this
case, Poisson distribution can be used effectively to model the
generation of data packets \cite{Rai_Lifetime_Modeling_DATE}. When
the average rate of packet generation, $\lambda$, is known, the
distribution of the number of data packets, $M$, generated by each
node, from time 0 to $T$ is
\begin{equation} \label{Packet generation model Poisson}
P(M=m) = \frac{e^{-\lambda T} (\lambda T)^m}{m!}
\end{equation}
where $m$ is a nonnegative integer number. Since the packet
generation distribution obeys the Poisson model, the time duration
between two consequent packet transmissions, $t$, has an exponential
distribution with mean $\frac{1}{\lambda}$:
\begin{equation} \label{Packet generation model Exp}
f_t(x)= \lambda e^{-x\lambda} u(x)
\end{equation}
where $u(x)$ denotes the unit step function.

We will consider the Poisson model for sensor's traffic in this
study. However, the proposed method can be extended to other traffic
distributions and data gathering scenarios.
\section{Lifetime Analysis in Single-hop Networks}\label{Section Prob Model Single Hop}
In this section, we derive the pdf of the lifetime in single-hop
WSNs. Assuming that nodes directly communicate with the sink, we
first derive the ccdf of the lifetime. Then, the pdf of the lifetime
is obtained by taking the derivative of the ccdf. The results are
extended to the case of multi-hop networks in Section \ref{Section
lifetime pdf multi-hop}.

It is assumed here that all of the nodes have the same initial
energy, same distribution over the area and the same packet
generation model. Other cases like nonuniform energy distribution or
different packet generation models are studied in Section
\ref{Section Extnesion}.

For the ease of presentation, the list of parameters is provided in
Table \ref{Parameters}. As mentioned, the lifetime of a single-hop
WSN is considered as the time when the ratio of dead nodes to the
total number of nodes, $N$, passes a threshold, $\beta$.

\begin{table}[!h]
\begin{center}
\begin{tabular}{|l l|}
\hline

$N$& Number of deployed nodes in the area\\
\hline

$\beta$& Threshold for the ratio of dead nodes to all nodes\\
\hline

$\alpha$& Path loss exponent\\
\hline

$k$& Path loss coefficient\\
\hline

$c$& Overhead energy\\
\hline

$E_i$& Initial energy in sensors\\
\hline

$\tau$ & Lifetime threshold\\
\hline

$t_i$& Lifetime achieved by sensor $i$\\
\hline

$L$& Lifetime achieved by the network\\
\hline

$\lambda$& Average rate of packet generation\\
\hline
$d_i$& Distance of sensor $i$ to the next hop\\
\hline
\end{tabular}
\caption{Parameters of the problem} \label{Parameters}
\end{center}
\end{table}
We start the network lifetime analysis by considering the lifetime
of one sensor. Defining $p_i$ as
\begin{equation}\label{Pi related to E}
p_i = \frac{E_i}{e(d_i)}
\end{equation}
for sensor $i$, it is clear that the maximum number of packets that
can be transmitted by this sensor is equal to $\lfloor p_i \rfloor$.

\lem \label{Lemma Sensor lifetime} If a sensor node with initial
energy $E_i$ is randomly placed in the area $\cal{R}$, the
probability of achieving a lifetime more than a threshold $\tau$
will be
\begin{equation}\label{ccdf of sensor lifetime}
P(t_i \geq \tau) = 1 - \frac{\gamma (\lfloor p_i \rfloor,\lambda
\tau)}{\Gamma(\lfloor p_i \rfloor)}
\end{equation}
where $\gamma( \cdot,\cdot)$ denotes the lower incomplete gamma
function
\begin{equation} \label{Lower Inc Gamma}
\gamma(a,x) = \int_0^x t^{a - 1} e^{-t}\,dt
\end{equation}
and $\Gamma(\cdot)$ represents the gamma function
\begin{equation}
\Gamma(x) = \int_0^{\infty} t^{x - 1} e^{-t}\,dt.
\end{equation}

\textit{Proof:} The lifetime of sensor $i$, $t_i$, depends on the
maximum number of packets that can be transmitted by the sensor to
the sink. Since $t_i$ is the sum of time durations between packet
transmissions until the last packet is sent by the sensor, we have
\begin{equation}\label{t_i the sum of t_ij}
t_i = \sum_{j=1}^{\lfloor p_i \rfloor} t_{ij}
\end{equation}
where $t_{ij}$ denotes the time duration between transmitting
packets $j-1$ and $j$ by sensor $i$, and $t_{i1}$ is defined as the
time when the first packet is transmitted. Since a Poisson model is
assumed for data packet generation, $t_{ij}$'s obey an exponential
distribution indicated in (\ref{Packet generation model Exp}). On
the other hand, it is known that the sum of independent identically
distributed (i.i.d) exponential random variables has a gamma
distribution \cite{Ross_Prob}. It is worthy to note that since the
node is deployed randomly in the area, the distance between the node
and the sink and consequently $p_i$ are a random variables. Hence,
given $p_i$, the conditional pdf of $t_i$ can be written as follows
\begin{equation}\label{pdf of t_i ineteger}
f_{t_i \vert p_i}(x) = \lambda ^{\lfloor p_i \rfloor } \frac
{x^{\lfloor p_i \rfloor - 1} e^{-\lambda x}} {\Gamma \left (\lfloor
p_i \rfloor \right )}  \quad x \geq 0.
\end{equation}
Now
\begin{align}
P(t_i \geq \tau \vert p_i) & =  1 - \int_0^{\tau} \lambda ^{\lfloor
p_i \rfloor} \frac {x^{\lfloor p_i \rfloor
-1} e^{-\lambda x}} {\Gamma( \lfloor p_i \rfloor)}\,dx \nonumber \\
& = 1 - \frac{\gamma (\lfloor p_i \rfloor,\lambda
\tau)}{\Gamma(\lfloor p_i \rfloor)} \label{ccdf of sensor lifetime}.
\end{align} \hfill $\blacksquare$

\prop \label{approx floorP by P} Since the fractional part of $p_i$
is usually much smaller than the integer part, $\lfloor p_i \rfloor
\simeq p_i$ and hence (\ref{ccdf of sensor lifetime}) can be
rewritten as
\begin{equation}\label{ccdf of sensor lifetime real}
P(t_i \geq \tau \vert p_i) = 1 - \frac{\gamma (p_i,\lambda
\tau)}{\Gamma(p_i)}
\end{equation}
For simplicity, we use (\ref{ccdf of sensor lifetime
real}) to analyze the network lifetime in the sequel.
$\square$

\corol \label{Corollary fixed power transmission} In the case of the
fixed transmission range, $r$, each node lives more than the
threshold with probability
\begin{equation}\label{Probability fixed transmission}
P(t_i \geq \tau)= 1 - \frac{\gamma (p_f,\lambda \tau)}{\Gamma(p_f)}
\end{equation}
where
\begin{equation}
p_f = \frac{E_i}{k r ^{\alpha} + c}.
\end{equation}
\textit{Proof:} In this case, all of the $p_i$'s have a
deterministic value equal to $p_f$. Therefore, the value of
$P(t_i \geq \tau)$ in (\ref{ccdf of sensor lifetime real}) is unconditional and
the proof is completed by replacing $p_i$ by $p_f$ in (\ref{ccdf of sensor lifetime real}).\hfill $ \square$

One can take another approach and approximate the value of $P(t_i
\geq \tau)$ to find a simpler form of (\ref{ccdf of sensor lifetime
real}).

\prop \label{Propositon Q function} Since $t_i$ in (\ref{t_i the sum of t_ij}) is the sum of i.i.d.
random variables, central limit theorem (CLT)  \cite{Ross_Prob}
indicates that its pdf tends to Gaussian distribution with mean
$\lfloor p_i \rfloor \lambda^{-1}$ and variance $\lfloor p_i \rfloor
\lambda ^{-2}$. Considering $\lfloor p_i \rfloor \approx p_i$, we
have
\begin{equation} \label{s_i Q_function}
P(t_i \geq \tau \vert p_i) = Q\left(\frac{\tau -
p_i\lambda^{-1}}{\sqrt{p_i}\lambda^{-1}}\right).
\end{equation}
where $Q(\cdot)$ is the ccdf of the normal distribution. \hfill $
\square$

To study the lifetime of the network, we consider the lifetime of
all the nodes in the network which necessitates the knowledge of
$p_i$ for all of the nodes in the network. When a node is deployed
randomly over an area, $p_i$ is a random variable with pdf
$f_{p_i}(x)$. In a random network deployment, $p_i$'s are usually
i.i.d. random variables and consequently have the same distribution,
$f_{p}(x)$. This distribution depends on the shape of the area,
energy dissipation model and the pdf of node distribution over the
area. In the Appendix, $f_p(x)$ is derived for some common area
shapes assuming a uniform distribution for the node deployment.

\theo\label{Lifetime theorem} Assuming $N$ equal-energy nodes are
distributed independently over the area $\cal{R}$, the probability
that the network achieves a lifetime more than a given threshold,
$\tau$, is equal to
\begin{equation}\label{Probability Formula}
P(L\geq \tau) = Q \left(\sqrt{N}\frac{1- \beta - \mu}{\sigma}
\right)
\end{equation}
where
\begin{align}\label{Mu Integral relation}
&\mu = \int_{\mathcal{R}} \left (1 - \frac{\gamma (x,\lambda
\tau)}{\Gamma(x)}\right) f_{p}(x)\,dx \\
\label{Mu and Sigma relation} &\sigma = \sqrt{\mu - \mu^2}.
\end{align}

\textit{Proof}: To find the number of nodes that live more than the
lifetime threshold, we define a Bernoulli random variable $l_i$
indicating the success of achieving the lifetime threshold by sensor
$i$:
\begin{equation}\label{Bernoulli RV}
l_i= \left\{
\begin{array}{ll}
1& \text{With probability equal to $s_i$},\\
0& \text{With probability equal to $1- s_i$}.
\end{array} \right.
\end{equation}
The success probability of $l_i$, given $p_i$, is equal to
\begin{equation}\label{s_i related to gamma function}
s_i = P(t_i \geq \tau \vert p_i) = 1 - \frac{\gamma (p_i,\lambda
\tau)}{\Gamma(p_i)}
\end{equation}
which was derived in Lemma \ref{Lemma Sensor lifetime}. The number
of live nodes after time $\tau$ can be found by defining a new
random variable, $w$, that denotes the number of successes in the
Bernoulli trials shown by $l_i$'s
\begin{equation}\label{W sum of Ki}
w = \sum_{i=1}^N l_i.
\end{equation}
Since nodes packet generations are independent and $p_i$'s are
i.i.d., $s_i$'s and consequently $l_i$'s are also i.i.d random
variables. 
In this case, $w$ has a binomial distribution
\cite{Nedleman_Bernoulli_Statistician}. Also, when the number of
trials is large enough, one can approximate the binomial
distribution with a Gaussian distribution. Since the number of nodes
are usually large enough, CLT can be applied on (\ref{W sum of Ki}).
Hence
\begin{equation}\label{Dist of W}
f_w(x)= \frac{1}{\sqrt{2\pi}\sigma_w}
\exp{-\frac{(x-\mu_w)^2}{2\sigma_w^2}}
\end{equation}
where $\mu_w$ is the mean and $\sigma_w^2$ denotes the variance of
$w$. From (\ref{W sum of Ki}), it is clear that
\begin{equation}\label{Mean of W}
\mu_w = \sum_{i=1}^N \mu_{l_i}
\end{equation}
where $\mu_{l_i}$ is the mean of $l_i$. Since $l_i$'s are
independent random variables
\begin{equation}\label{Var of W}
\sigma_w^2 = \sum_{i=1}^N \sigma_{l_i}^2
\end{equation}
where $\sigma_{l_i}^2$ is the variance of $l_i$. To find the values
of $\mu_w$ and $\sigma_w$, we need to have the unconditional mean
and variance of $l_i$'s using the conditional values. Since $l_i$'s
are Bernoulli random variables
\begin{equation}\label{Mean and var k_i Cond}
\mu_{l_i \vert p_i}= s_i, \quad \sigma_{l_i \vert p_i}^2 = s_i -
s_i^2.
\end{equation}
On the other hand, for two random variables $x$ and $z$, the
unconditional mean and variance of $x$ can be found using the
conditional mean and variance as follows \cite{Ross_Prob}
\begin{align}
&\mu_x = E[\mu_{x \vert z}] \label{mean conditional} \\
&\sigma^2_x = E[\sigma^2_{x \vert z}] + \text{Var}[\mu_{x \vert z}]
\label{Var conditional}
\end{align}
where $E[\cdot]$ is the expected value and $\text{Var}[\cdot]$
denotes the variance of the random variable. Using (\ref{s_i related
to gamma function}), (\ref{Mean and var k_i Cond}), (\ref{mean
conditional}) and (\ref{Var conditional}), it can be shown that
\begin{align}
&\mu_{l_i} = E[s_i] = \int_{\mathcal{R}} \left (1 - \frac{\gamma
(x,\lambda
\tau)}{\Gamma(x)}\right) f_{p_i}(x)\,dx\\
&\sigma^2_{l_i} = E[s_i - s_i^2] + \text{Var}[s_i] = E[s_i] -
E^2[s_i] = \mu_{l_i} - \mu^2_{l_i}.
\end{align}
Since $p_i$'s are i.i.d random variables with pdf $f_p(x)$, we have
\begin{align}
&\mu_{l_i} = \mu = \int_{\mathcal{R}} \left (1 - \frac{\gamma
(x,\lambda \tau)}{\Gamma(x)}\right) f_{p}(x)\,dx   \quad \forall i \\
&\sigma_{l_i} = \sigma = \sqrt{\mu - \mu^2} \quad \forall i.
\end{align}
Then, using (\ref{Mean of W}) and (\ref{Var of W})
\begin{equation}\label{mu and var w }
\mu_w = N\mu, \quad \sigma^2_w = N\sigma^2
\end{equation}

To derive the probability of achieving the lifetime threshold by the
network, we just need to know the probability of achieving the
lifetime by at least $(1 - \beta)N$ nodes. Hence
\begin{align}
F_L^c(\tau) = P(L\geq \tau) & = P(w \geq (1-\beta)N)\nonumber\\
&= Q\left(\sqrt{N}\frac{1- \beta - \mu}{\sigma} \right)
\end{align}
where $F_L^c(\tau)$ represents the ccdf of the network lifetime.
\hfill $\blacksquare$

\prop \label{Proposition mu Q function} Using Proposition
\ref{Propositon Q function}, $\mu$ can also be calculated as
\begin{equation}
\mu = \int_{\mathcal{R}} Q\left(\frac{T_{\mathrm{thr}} -
x\lambda^{-1}}{\sqrt{x}\lambda ^{-1}}\right) f_{p}(x)\,dx.
\end{equation}\hfill $ \square$

\corol \label{Corollary lifetime PDF} Assuming a network with
parameters given in Theorem \ref{Lifetime theorem}, the probability
distribution function of the network lifetime is
\begin{align}\label{Lifetime pdf}
f_L(\tau)=&\frac{\lambda\sqrt{N}}{2\sqrt{2\pi}} \frac{1-\mu-\beta( 1
- 2\mu) }{(\mu - \mu^2)^{\frac{3}{2}}} c(\tau) e^{-(\lambda \tau
+\frac{N(1-\beta -\mu)^2}{2(\mu - \mu^2)})} \nonumber \\
& 0\leq \tau \leq \infty
\end{align}
where
\begin{equation}
c(\tau) = \int_{\mathcal{R}} \frac{f_p(x)}{\Gamma(x)}(\lambda
\tau)^{x-1}\,dx.
\end{equation}
\textit{Proof}: The ccdf of the network lifetime was derived in the
previous theorem. Then we have
\begin{equation}
f_L(\tau)=
-\frac{d(F^c_L(\tau))}{d\tau}=-\frac{d(F^c_L(\mu))}{d\mu}\frac{d\mu}{d\tau}
\end{equation}
which results in (\ref{Lifetime pdf}). \hfill $\blacksquare$

\section{Lifetime Analysis in Multi-hop Networks} \label{Section lifetime pdf multi-hop}
In multi-hop networks, the network lifetime depends on the way that
the routing scheme distributes the traffic load among the sensor
nodes. The minimum cost routing (minimum required energy or minimum
number of hops) is conventionally used in wireless networks.
However, this routing scheme cannot guarantee the maximum lifetime
in the network \cite{Al-Karaki_Routing_Magazine2004}. On the other
hand, maximum lifetime routing attempts to prolong the network
lifetime by proper traffic distribution among the nodes. This scheme
may not have the minimum overall consumed energy. Since we mainly
focus on the lifetime analysis, we just consider the maximum
lifetime routing. Nevertheless, the proposed approach can be used for other routing schemes knowing how the traffic is distributed among the nodes.

In multi-hop networks, the whole network traffic passes through the
nodes in the vicinity of the sink, hence, death of these nodes can
have a significant effect on the network performance. Therefore, we
need to modify our previous definition of the lifetime.

Assume that $\mathcal{H}$ shows the set of nodes that are in the
vicinity of the sink and directly communicate with it. Since all
other nodes communicate to the sink through these nodes, they will
be out of energy sooner than the other ones. So, we define the
lifetime based on the ratio of dead nodes within $\cal{H}$ to $\vert
\mathcal{H} \vert$ where $\vert \cdot \vert$ denotes the cardinality
of the set. It is also assumed that the sensors are distributed over
a circle with radius $R$ and the data sink is positioned at the
center of the area. The lifetime of the network over other area
shapes will be discussed later.

We assume that the sensors perform transmission with a fixed power
which results in a fixed transmission range $r$. Based on the
maximum transmission radius of the sensors and the area radius, the
area can be divided to a number of rings (Figure \ref{Fig Tx
rings}). The sensors within a ring send their data to the sensors
within the neighboring inner ring. The number of rings, $n$, within
the area can be simply found as
\begin{align} \label{Number of tiers}
n = \left \lceil \frac{R}{r} \right \rceil
\end{align}
where $\lceil \cdot \rceil$ denotes the integer ceiling. For
simplicity, it is assumed that $R$ is an integer multiple of $r$.
This assumption allows us to focus on methodologies and can be
removed if necessary. In addition, since each ring carries the
traffic of all outer rings, the average traffic carried by the
sensors within each ring is different and depends on the distance of
the ring to the sink. To study the network lifetime, we consider the
case when the routing scheme distributes the network traffic equally
between the nodes within each ring. This scheme prevents the nodes
from being exhausted quickly and prolongs the lifetime.

\begin{figure}[!h]
\centering
\includegraphics[scale=.11]{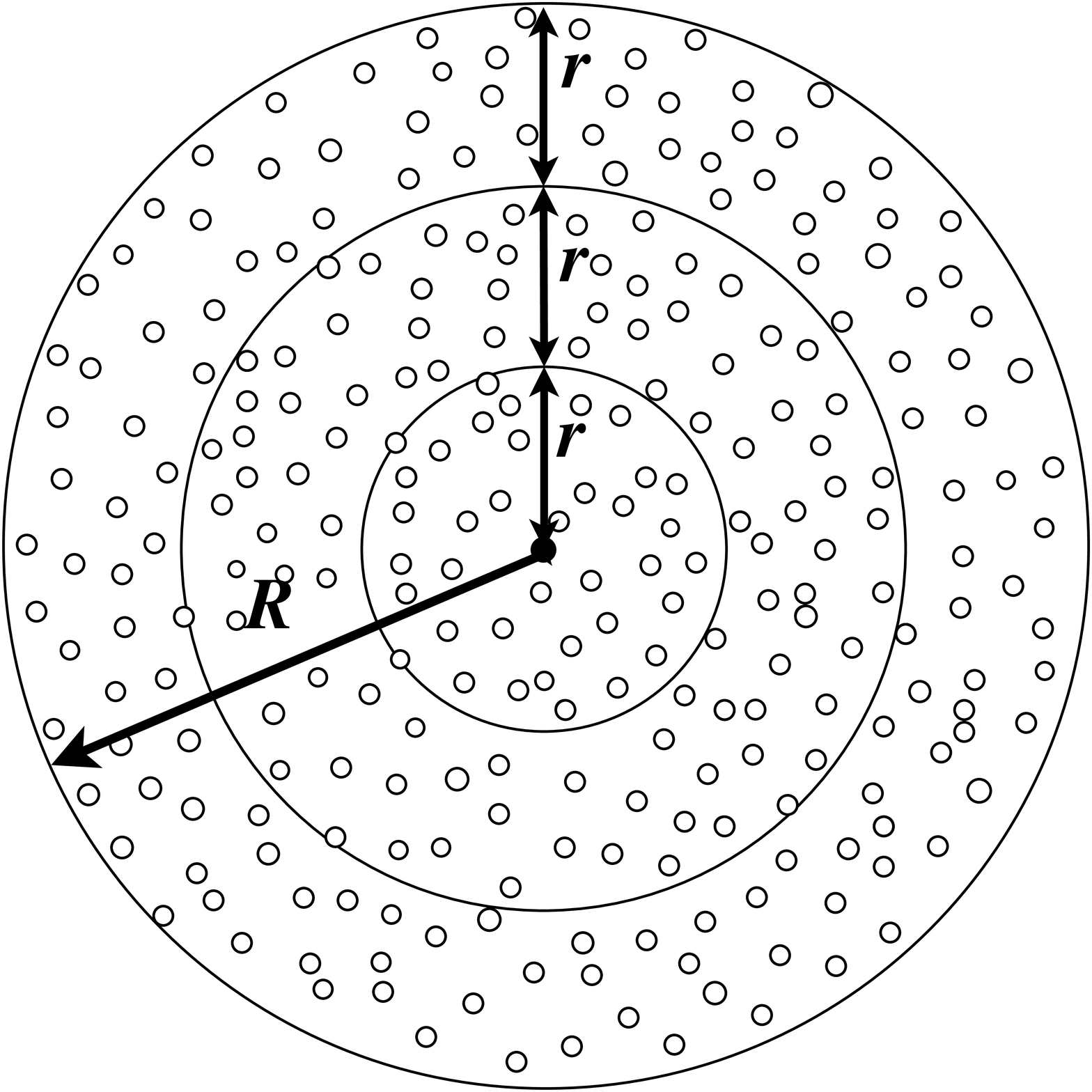}
\caption{Rings within a multi-hop network} \label{Fig Tx rings}
\end{figure}

Based on the assumed routing scheme, the average rate of the packet
transmission by each node within ring $i$ is equal to
\begin{equation}\label{lambda of tiers}
\lambda_i = \lambda \frac{N - \sum_{j=1}^{i-1} N_j}{N_i} \quad
\forall i = 1,2,\ldots,n
\end{equation}
where $N_i$ denotes the number of sensors within ring $i$. Since
nodes are assumed to be deployed randomly in the area, $N_i$ is a
binomial random variable. If one assumes a uniform deployment for
the nodes, $N_i$ will have a binomial distribution with mean $Nq_i$
where
\begin{equation}\label{binomial chance for rings}
q_i = \frac{r^2 (2i - 1)}{R^2}
\end{equation}
represents the probability of positioning a sensor in the ring $i$.
Therefore, the time duration between two consequent transmissions,
$t$, by a node in the ring $i$ obeys an exponential distribution as
follows
\begin{equation}\label{exponential model for rings}
f_{t \vert N_i}(x) = \lambda_i e^{-x \lambda_i} u(x).
\end{equation}
Since the lifetime is mainly effected by the nodes within the first
tier, we just consider the probability of achieving the lifetime
threshold by the first ring. Nevertheless, the probability of
achieving $\tau$ by other rings can also be investigated using
(\ref{exponential model for rings}). As discussed earlier,
probability of achieving a lifetime threshold depends on the number
of nodes within the area. Hence, using Theorem \ref{Lifetime
theorem} and Corollary \ref{Corollary fixed power transmission}, one
can find the conditional probability of achieving $\tau$ by the
first ring
\begin{equation}\label{ccdf for the first ring}
P(L\geq \tau \vert N_1) = Q \left(\sqrt{N_1} \frac{1 - \beta -
\mu}{\sigma}\right)
\end{equation}
where
\begin{align}
&\mu = 1 - \frac{\gamma(p_f,\lambda_1\tau)}{\Gamma(p_f)} \\
&\sigma = \sqrt{\mu - \mu^2}.
\end{align}
Therefore, by removing the condition on $N_1$ in (\ref{ccdf for the
first ring}), we have
\begin{equation}\label{Unconditioing Ni}
P(L\geq \tau) = \sum_{j=0}^N P(L\geq \tau \vert N_1) P(N_1 = j)\quad
n_1 = 1,2,\ldots,N
\end{equation}
where
\begin{equation}
P(N_1 = n_1) = \nchoosek{N}{n_1} q_1^{n_1} (1- q_1)^{N - n_1}.
\end{equation}

The given discussion is not restricted to circular areas and can
also be applied to other area shapes. To study the lifetime of the
network in other area shapes, we just need to recalculate the value
of $q_1$ as follows
\begin{equation}\label{q1 other area shapes}
q_1 = \frac{\pi r^2}{S}
\end{equation}
where $S$ is the size of area. Then, ccdf of the lifetime is derived
by putting this value of $q_1$ into (\ref{Unconditioing Ni}).

\section{Some Notes} \label{Section Extnesion}
In Section \ref{Section Prob Model Single Hop}, we considered the
finite number of nodes in the area. We will study the asymptotic
analysis in this section. Also, we earlier studied the case when all
of the sensors have the same features such as traffic model, initial
energy and deployment. In addition, the packet generation model was
supposed to be Poisson. Here, we provide some discussions on the
results in Section \ref{Section Prob Model Single Hop} and
generalize them for more cases.

\subsection{Asymptotic Analysis}
Since the lifetime ccdf in (\ref{Probability Formula}) depends on
the number of nodes distributed over the area, we can study the
effect of the node density on the probability of achieving the
lifetime threshold.

\corol \label{Corollary on node density} The probability of
achieving a lifetime threshold approaches 0 or 1 by increasing the
number of nodes.

\textit{Proof:} For large $N$, two cases can happen depending on the
sign of $a = 1 - \beta - \mu$. Since $Q$-function is a decreasing
function, when $a
> 0$, increasing $N$ causes the probability of hitting the lifetime
threshold to tend $Q(\infty) = 0$. In other words, almost surely the
given lifetime threshold cannot be achieved. Now, considering that
\cite{Salehi_Commsystems}
\begin{equation}
\frac{1}{\sqrt{2\pi}x}\left(1 - \frac{1}{x^2}\right)
e^{-\frac{x^2}{2}} < Q(x) < \frac{1}{\sqrt{2\pi}x}
e^{-\frac{x^2}{2}} \quad \forall x\geq  0
\end{equation}
the rate of the probability decay is proportional to $e^{-N}$. In a
similar manner, the probability approaches $Q(-\infty) = 1$ when $a
< 0$. That is, the network almost surely achieves the lifetime
threshold. The error in this prediction also decays exponentially
with $N$. \hfill $\blacksquare$

An interesting case occurs when one considers the lifetime of the
network based on the death of the first node. In this case $\beta =
\frac{1}{N}$, which approaches 0 when $N$ increases. Hence,
according to the Corollary \ref{Corollary on node density}, it is
necessary to consider just the sign of $1 - \mu$ in order to predict
the asymptotic behavior of the network lifetime (i.e. when
$N\rightarrow \infty$). Assuming $\tau > 0$, we have
\begin{equation}
\mu = \int_{\mathcal{R}} \left (1 - \frac{\gamma (x,\lambda
\tau)}{\Gamma(x)}\right) f_{p}(x)\,dx < \int_{\mathcal{R}}
f_{p}(x)\,dx = 1
\end{equation}
and consequently $1 - \mu > 0$. Therefore, under this stringent
definition of the lifetime, the probability of achieving the
lifetime $\tau$ approaches 0 as $N$ increases.

\subsection{Different Traffic Models}
In Section \ref{Section Prob Model Single Hop}, we considered the
case when all of the sensors have the same Poisson model for the
packet generation. Here, we consider two other cases: 1) The average
rate of packet generation changes with the position of the sensor,
2) packet generation obeys another model rather than Poisson. It is
worthy to note that the assumed model is similar for all of the
sensors.

If the average rate of packet generation, $\lambda$, varies with the
position of the sensor (e.g. due to the spatial correlation of data
or data aggregation and compression), we have the mean and variance
of $l_i$ conditioned on both $p$ and $\lambda$. To derive the
unconditional mean and variance of $l_i$, we need to calculate
\begin{equation}\label{joint lambda P}
\mu_{l_i} = \mu = \int \int_{\mathcal{R}} \left (1 - \frac{\gamma
(x,\lambda
y)}{\Gamma(x)}\right) f_{p,\lambda}(x,y)\,dx dy\\
\end{equation}
where $f_{p,\lambda}(x,y)$ denotes the joint pdf of $p$ and
$\lambda$. Other parts of the analysis will remain unchanged.

Also, the proof given for Theorem \ref{Lifetime theorem} can be
applied to the cases when the traffic model obeys another pattern
rather than Poisson model. Assume that the pdf of the time duration
between two packet transmissions follows a model with mean $\mu_t$
and variance $\sigma_t^2$. Using CLT, $t_i$ can be accurately
approximated by a Gaussian distribution with mean $p_i \mu_t$ and
variance $p_i \sigma_t^2$. The remaining part of the proof is
unchanged.

The proposed analysis can also be extended to time-driven networks.
In this case, the time duration between two consequent transmissions
is fixed and is equal to $T$. Hence
\begin{equation}\label{sensor lifetime time driven}
t_i = \lfloor p_i \rfloor T.
\end{equation}
The unconditional values of $\mu$ and $\sigma$ is found by
integration over $p_i$. Then, the result given in Theorem
\ref{Lifetime theorem} can be applied.

\subsection{Nonuniform Energy Distribution}
Assume that the energy is distributed over the network in a
nonuniform way. As a consequence, $s_i$'s in (\ref{s_i related to
gamma function}) are not identically distributed. This may also
arise when the sensors generate packets with different rates (i.e.
nonidentical Poisson distributions). In this situation, $w$ does not
have any standard distribution, however, we can still use CLT to
approximate the pdf of $w$ with a Gaussian distribution. To this
end, we will give a brief discussion on the probability of achieving
the lifetime threshold by the network.
\lem\label{Lemma Surprising} Assume that $z_i$'s ($1\leq i \leq m$)
are $m$ independent random variables such that
\begin{equation}
\sum_{i=1}^m \mu_{z_i} = m \overline{\mu}.
\end{equation}
where $\mu_{z_i}$ denotes the mean of $z_i$. Also, $X_i$'s ($1\leq i
\leq m$) are $m$ Bernoulli trials such that
\begin{equation}
P(X_i = 1) = z_i \quad \forall i.
\end{equation}
Now, if $X$ denotes the sum of $X_i$'s, the variance of $X$ is
maximum when
\begin{equation}
\mu_{z_i} = \overline{\mu}\quad \forall i=1,\ldots,m
\end{equation}
(see \cite{Nedleman_Bernoulli_Statistician} for the proof).

\corol \label{Corollary} For nonidentical distributed $s_i$'s such
that
\begin{equation}
\mu_w = \sum_{i=1}^N \mu_{l_{i}}=\sum_{i=1}^N E[s_i]=N \mu
\end{equation}
(\ref{Probability Formula}) is an upper bound for the probability of
achieving the lifetime when $1 - \beta - \mu >0$, otherwise it is a
lower bound.

\textit{Proof:} Since we assumed the identical distribution in the
proof of Theorem \ref{Lifetime theorem}, Lemma \ref{Lemma
Surprising} indicates that $\sigma_w$ in (\ref{mu and var w }) is
the maximum possible variance of $w$. The proof is completed
considering the decreasing property of the $Q$-function. \hfill
$\blacksquare$

\section{Experimental results}\label{Section Simulations}
In this section, we investigate the accuracy of the proposed
analysis through some experiments. We first study the probability of
achieving a lifetime threshold in single-hop networks. To this end,
the simulations are performed over different area shapes with the
same area size to investigate the effect of the area shape. In
addition, the effect of the node density is studied. Moreover,
simulations are performed to study the network lifetime in multi-hop
networks. Through these simulations, it will be shown that how the
transmission range and consequently the number of hops effect the
lifetime of the network.

\subsection{Single-hop Networks}
The parameters of model (\ref{Power dissipation model}) depend on
the data rate, antenna height, antenna gain, etc. Typical values of
$e_t$ and $e_o$ are given in \cite{Heinzelman_Phd_Thes}. For $\alpha
= 4$, which we use in our simulations, the values of $e_t$ and $e_o$
are respectively 0.0013 pJ/bit/$\text{m}^4$ and 50 nJ/bit for a
1Mbps data stream. Here, It is assumed that the packets have 1000
bits length, hence, $k =$1.3 pJ/$\text{m}^4$ and $c =$50 $\mu$J in
(\ref{Power dissipation model}).

Network has 500 nodes that are deployed uniformly and sink is
positioned at the center of the area. Also, the packet generation
model obeys the Poisson distribution and each sensor sends its
packets directly to the sink. All of the sensors have the same
initial energy equal to 11 mJ. Assuming that sensors send packet
with the average rate of 1 packet/hour, the probability of achieving
the lifetime of 100 hours is studied through simulation.

To investigate the effect of the area shape on the lifetime, the
simulations are carried out over areas with the same size equal to
$100 \pi \text{m}^2$ but with different shapes. To decrease the
final result variance and reach the proper confidence interval, the
simulation is run 10000 times over each area and the results are
averaged.

\begin{figure}[!h]
\centering
\includegraphics[scale=.41]{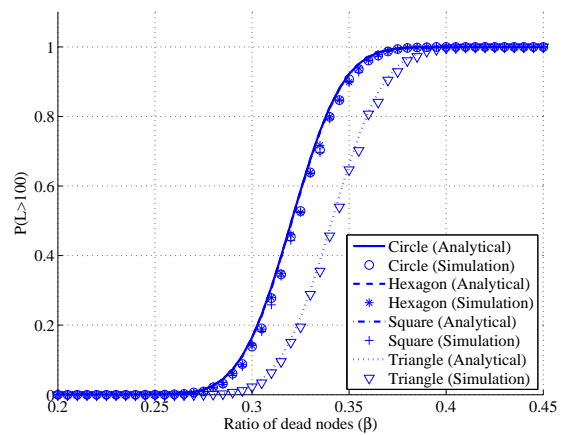}
\caption{Probability of achieving the lifetime threshold vs. the
ratio of dead nodes for single-hop networks deployed over different
area shapes} \label{Beta Diff area shapes}
\end{figure}

Figure \ref{Beta Diff area shapes} depicts the probability of
achieving the lifetime threshold vs. the ratio of dead nodes over
circular, hexagonal, squared and triangular areas. As it can be
seen, the probability of achieving the lifetime threshold in
circular, hexagonal and squared areas are very close. Since in a
triangle, the distance of the sensors to the sink is more
non-uniform and it has the largest circumcircle compared to other
area shapes, triangle has a smaller probability to achieve the
lifetime threshold.

As discussed through the paper, depending on the value of $1 - \mu -
\beta$ and by increasing the number of nodes, it can be almost
surely determined whether the network achieves a lifetime threshold
or not. The effect of the node density on achieving the lifetime
threshold is shown in Figure \ref{Fig Node density effect}. The
lifetime of the network is considered as the moment when 0.3 of the
nodes in the network die. In the first case, $E_i = 11$ mJ which
results in $1 - \beta - \mu > 0$. Hence, as discussed in Section
\ref{Section Extnesion}, the desired probability decreases by
increasing $N$ which is verified by the simulation. In the second
case, the initial energy is equal to 11.6 mJ which causes $1 - \beta
- \mu < 0$. As shown in Figure \ref{Fig Node density effect},
the probability of achieving the desired lifetime is an increasing
function of $N$.

\begin{figure}[!h]
\centering
\includegraphics[scale=.41]{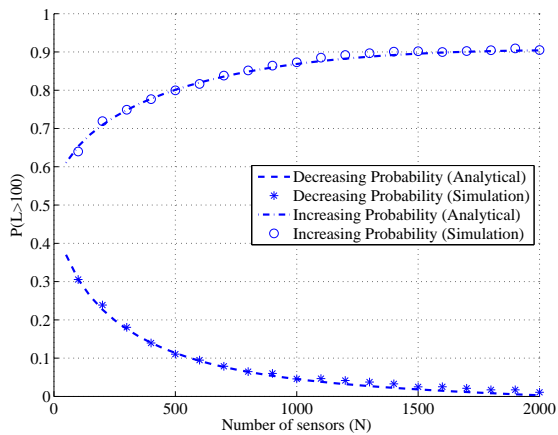}
\caption{Probability of achieving the lifetime threshold vs. the
number of sensors in a single-hop network} \label{Fig Node density
effect}
\end{figure}

\subsection{Multi-hop Networks}
To study the network lifetime in a multi-hop network, it is assumed
that 500 nodes are deployed uniformly over a circle with radius 100 m. All of the nodes have the same initial energy
equal to $E_i = 100$ mJ. The parameters in (\ref{Power dissipation
model}) are kept the same as the previous part. A greedy routing
algorithm is used to balance the network traffic such that data
packets are identically distributed between the nodes in the first
ring of the network, $\cal{H}$. Considering this fact that all of
the nodes use a constant transmission power and the traffic is
distributed identically between the first-ring nodes, all of the
nodes within $\cal{H}$ have approximately similar lifetime. As a
consequence, they die in time moments very close to each other.
Therefore, we can say that the desired probability is not
significantly effected by the value of $\beta$ (Figure \ref{Fig Beta
Multihop}).

\begin{figure}[!h]
\centering
\includegraphics[scale=.41]{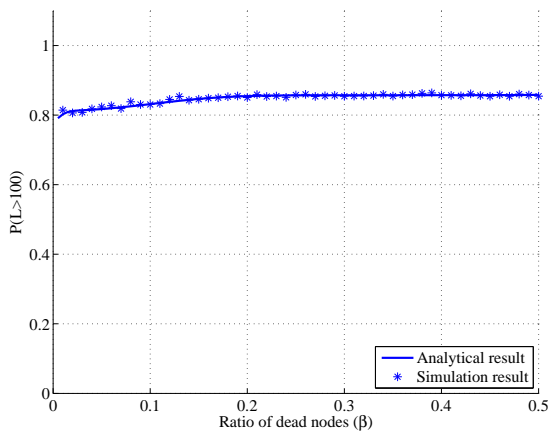}
\caption{Probability of achieving the lifetime threshold vs. the
ratio of dead nodes in a multi-hop network} \label{Fig Beta
Multihop}
\end{figure}

It is interesting to study the effect of the transmission range and
consequently the number of hops on the lifetime. Figure \ref{Fig
multihop Tx range} depicts the probability of reaching the lifetime
threshold vs. the transmission range. The lifetime is considered as
the moment when 0.3 nodes within $\cal{H}$ are dead. By
decreasing $r$, number of nodes within $\cal{H}$ decreases, hence,
they carry more packets and will die earlier. Therefore, it is
expected that the desired probability decreases by reducing $r$.
Indeed, while the nodes far from the sink still have enough energy
to send packets, the nodes within $\cal{H}$ cease. To overcome this
drawback, nonuniform energy distribution can be applied
\cite{IEEE_Monograph}. Also, the fixed transmission power causes the
nodes within $\cal{H}$ to die sooner compared to the case where nodes adjust their transmission power.

\begin{figure}[!h]
\centering
\includegraphics[scale=.41]{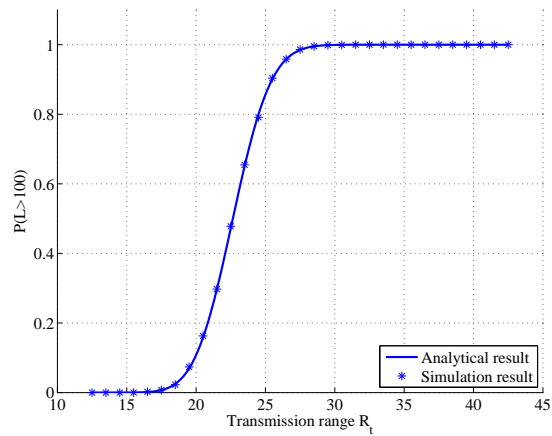}
\caption{Probability of achieving the lifetime threshold vs. the
transmission range in a multi-hop network} \label{Fig multihop Tx
range}
\end{figure}

\section{conclusions}\label{Section Conclusion}
In this paper, we considered the problem of finding the probability
of achieving a lifetime threshold by the network which is equivalent
to finding the ccdf of the network lifetime. Using the power
consumption model of (\ref{Power dissipation model}), the ccdf of
the lifetime was derived for the single-hop networks. To this end,
it was assumed that all of the nodes have identical packet
generation model, initial energy and random deployment in the area.
The methodology was also extended to the case when these conditions
may not be satisfied. Then, the problem was studied for the
multi-hop case. In addition, the asymptotic relation between the
number of nodes and the lifetime was investigated. Through some
simulations, the accuracy of our analysis was investigated for the
networks deployed over different area shapes. Using the proposed
method, one can design both node and network parameters (e.g. node
density, data rate, initial energy) according to the desired
lifetime.

\section*{Appendix}\label{Appendix pdf of p}
The pdf of the network lifetime depends on the distribution of the
maximum possible number of packet transmissions by each node, $p$.
In this appendix, we find the pdf of $p$ over some common area
shapes. The pdf of $p$ over a circle area is required for finding
the lifetime pdf of multi-hop networks in Section \ref{Section
lifetime pdf multi-hop}. Also, this pdf over regular polygons is
useful for studying the lifetime of a network composed of clusters
tiling the area.

\subsection{Network Deployed Over a Circle}
Assume that the nodes are deployed uniformly over a circle with radius $R$.
Also, assume that the sink is located at the center of the circle.
Since the nodes are deployed uniformly over the area, the pdf of the
distance between the nodes and sink, $d$, is
\begin{equation}
f_d(x) = \left\{
\begin{array}{ll}
\frac{2x}{R^2}  & \quad 0 < x \leq R\\
0 & \text{\small{Otherwise}}
\end{array}.
\right.
\end{equation}
Now, using the energy consumption model (\ref{Power dissipation
model}), we have the following expression for the pdf of $p$
\begin{align}\label{f_p circle}
f_{p}(x)  = \left\{
\begin{array}{ll}
\frac{2E_i}{R^2 k \alpha x^2} \left[\frac{E_i -
cx}{kx}\right]^{\frac{2-\alpha}{\alpha}}&\frac{E_i}{kR^{\alpha} +
c}\leq x < \frac{E_i}{c}\\
0 & \text{\small{Otherwise}}
\end{array}. \right.
\end{align}

\subsection{Network Deployed Over a Regular Polygon}
Suppose that the sensors are deployed over a regular polygon having
$n$ equal sides with length $a$. Again, we assume that the sink is
placed at the center of the area. In this case, we have
\begin{equation}
f_d(x) = \left\{
\begin{array}{ll}
\frac{2\pi x}{S} & 0 < x \leq r_i \\
\frac{2\pi x - 2 n x \cos^{-1}{\frac{r}{x}}}{S} & r_i < x \leq R_c\\
0 & \text{\small{Otherwise}}
\end{array} \right.
\end{equation}
where
\begin{equation}
r_i = \frac{a}{2} \cot{\frac{\pi}{n}}
\end{equation}
is the radius of the inscribed circle of the polygon,
\begin{equation}
R_c = \frac{a}{2 \sin{\frac{\pi}{n}}}
\end{equation}
represents the radius of the circumcircle of the polygon and
\begin{equation}
S = \frac{n}{4} a^2 \cot{\frac{\pi}{n}}
\end{equation}
denotes the polygon area. Now, using the relation between $d$ and
$p$, we have
\begin{align}
f_p(x) = \frac{2 E_i}{k \alpha S x^2} \left[
\frac{E_i-cx}{kx}\right]^{\frac{2-\alpha}{\alpha}} \left[\pi - n
\cos^{-1}{\frac{r}{\left (\frac{E_i -
cx}{kx}\right)^{\frac{1}{\alpha}}}} \right]
\end{align}
when
\begin{equation}
\frac{E_i}{kR_c^{\alpha} + c}\leq x < \frac{E_i}{kr_i^{\alpha} + c}
\nonumber
\end{equation}
and
\begin{equation}
f_p(x) = \frac{2 \pi E_i}{k \alpha S x^2}
\left[\frac{E_i-cx}{kx}\right]^{\frac{2-\alpha}{\alpha}}
\end{equation}
when
\begin{equation}
\frac{E_i}{kr_i^{\alpha} + c}\leq x < \frac{E_i}{c} \nonumber
\end{equation}
and $f_p(x) = 0$ elsewhere.


\end{document}